# Vibrational properties of *sp* carbon atomic wires in cluster–assembled carbon films

**Giovanni Onida**[*,1,2], **Nicola Manini**[1,2], **Luca Ravagnan**[1,4], **Eugenio Cinquanta**[3,4] **Davide Sangalli**[1,2] **Paolo Milani**[1,4]

[1] Dipartimento di Fisica dell'Università degli Studi di Milano, via Celoria 16, 20133 Milano, Italy
[2] European Theoretical Spectroscopy Facility (ETSF)
[3] Dipartimento di Scienza dei Materiali, Università di Milano Bicocca, Via Cozzi 53, 20125 Milano, Italy
[4] CIMAINA - Via Celoria 16, 20133 Milano, Italy

[*] Corresponding author: e-mail giovanni.onida@unimi.it, Phone: +39-02-50317407, Fax: +39-02-50317482

Linear chains made by a single row of *sp*-hybridized carbon are predicted to display fascinating mechano-electronic properties connected with their termination and stabilization inside realistic carbon structures. The present work describes how the computed vibrational properties of cumulenic and polyynic carbon chains allow one to interpret the carbynic features observed in Raman spectra of cluster-assembled $sp - sp^2$ films. The overall picture is consistent with the measured decay of the *sp* components induced by air or oxygen exposure.

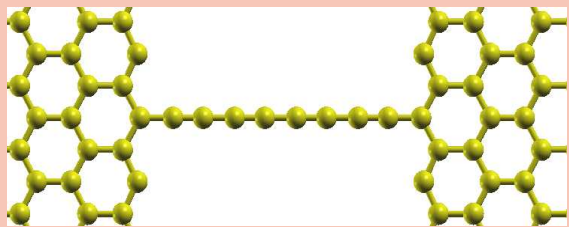

The electronic properties of carbon atomic chains attached to the edges of graphitic fragments can be controlled by a relative twist of the terminating planes.

## 1 Introduction

Fullerenes, nanotubes, and graphene are examples of the wide class of structures based on $sp^2$-hybridized carbon. Carbon atoms can also show an *sp* hybridization, and hence form *linear* structures. Linear atomic chains made by a single row of *sp*-hybridized carbons, known as Carbynes, are indeed the most one-dimensional among all conceivable carbon nanostructures. In the last years, the possibility to use them as conducting bridges between metallic leads, or even between $sp^2$ or $sp^3$ carbon fragments, has been suggested [1]. The basic assumption is a possible delocalization of $\pi$ electrons along the chain, allowing for charge transport.

However, as we show below, this one-dimensional $\pi$ electron delocalization differs substantially from that observed in aromatic compounds, based on $sp^2$-hybridized carbons.

Traditionally, finite carbynes are classified as polyynes or cumulenes, according to the nature of their bonding: the first ones display a sequence of single and triple bonds – and acquire a substantial bond length alternation– while the second ones have all double bonds, all of about the same length.

Until recently the "carbyne" scientific community had a small interaction/superposition with the community of people studying $sp^2$ carbon systems, such as fullerenes and graphene. Works suggesting the relevance of carbynic chains in graphene-based devices appeared very recently [2]; in particular, carbynes were suggested to form bridges across nanometric gaps in graphene. This conjecture was advanced in order to explain the reversible conductance switching observed in two-terminal graphene devices [2, 3]

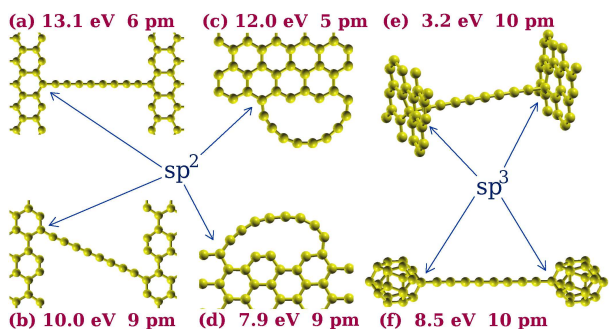

**(a) 13.1 eV  6 pm   (c) 12.0 eV  5 pm   (e) 3.2 eV  10 pm**

**(b) 10.0 eV  9 pm   (d) 7.9 eV  9 pm   (f) 8.5 eV  9 pm**

**Figure 1** A few representative structures involving an 8-atoms sp-bonded carbon chain terminated on $sp^2$ carbon fragments [(a-e): NRs; (f): $C_{20}$]. Either edge termination [(a-d), $sp^2$-like] or termination on an internal atom of the fragment [(e-f), $sp^3$-like] is possible. Binding energies (with respect to the uncapped straight chain plus fully relaxed $sp^2$ fragments) and BLA are reported.

On the other hand, the coexistence of $sp^2$ and $sp$ hybridization in Carbon has been demonstrated since 2002, based on Raman spectroscopy. In particular, the experimental group of Ref. [4,5] uses supersonic cluster beam deposition to produce nanostructured carbon films with a clear signature of sp-bonded carbon, as demonstrated by measuring in situ the Raman spectra.

The carbynic Raman peak, sometimes identified as "C-band", is due to the stretching of the $sp$ bond and appears around 2000 cm$^{-1}$, being well separated from the lower structures associated with the stretching of the $sp^3$ and $sp^2$ carbon-carbon bonds. The C-band, however, is found to decay with time and with exposure of the sample to oxygen, with the $sp$ component almost disappearing after 3 days [6]. More recent low-noise, high-statistics experiments have allowed one to disentangle, from a purely empirical analysis of the experimental data, the presence of at least five different components building up this carbynic peak, each with a different decay rate [6]. Moreover, 4 out of 5 components also display a blueshift and a narrowing during the decay. Another relevant experimental finding is the reduction of the electrical conductance of the sample associated with the decay of the carbynic Raman peak.

**2 Theoretical methods** In order to gain more insight in the physics of such $sp + sp^2$ systems, we performed a set of *ab-initio* total energy calculations [1]. We carried out *ab-initio* calculations within the Density-Functional theory in the local-density approximation (DFT-LDA) with a plane-waves package [8] using default ultrasoft pseudopotentials, and wavefunction/charge cutoffs of 15/120 Hartree. We relax the atomic positions until the largest residual force is less than $10^{-4}$ Hartree/$a_0$ (8 pN). We compute the vibrational properties within density-functional perturbation theory [9]. Non-resonant Raman intensities were obtained by numerically evaluating second derivatives of the atomic forces with respect to the electric field [10].

**3 Results** In the following we present theoretical and experimental results. The calculations are aimed i) to identify the most stable structures and geometries; and ii) to compute the Raman frequencies and intensities for a selection of realistic structures. Computed Raman spectra are compared with measured spectra of cluster-assembled $sp - sp^2$ carbon films.

**3.1 Structure and energetics** First of all, we notice that there are essentially two ways to terminate a linear chain on an $sp^2$ graphitic fragment: either the chain connects perpendicularly to the surface of the fragment (or the surface of a fullerenic cage), by means of an "interface" atom having the $sp^3$ hybridization, like in the examples shown in Fig. 1e-f, or the chain can be attached to an edge, like in Fig. 1a-d. In the latter case the "interface" atom is also $sp^2$. The first remarkable result of total energy calculations is the large binding energy which is obtained in this second case, especially for the attachment to a zigzag edge.

Figure 2 shows the result of a geometry optimization for a linear chain put in the vicinity of the edge of a graphene fragment or nanoribbon, ending in different local minima according to the initial conditions. For an 8-atoms long chain, if the initial distance is large enough the chain bends, and attaches its two ends to the edge (Fig. 2); if the distance is smaller, the chain makes multiple bonds and the nanoribbon grows, possibly incorporating defects. Remarkably, the energy gain *per new bond* is much larger in the first case than in the second.

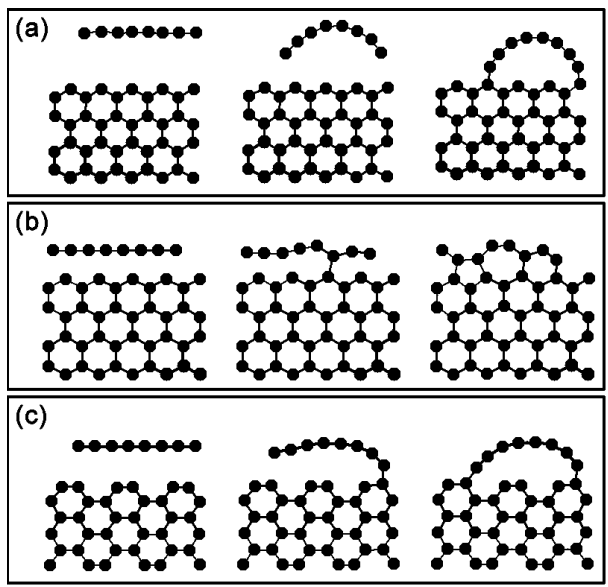

**Figure 2** Successive minimization steps for the attachment of a $C_8$ chain to nanoribbon edges. (a): the chain starts 2.1 Å from the zigzag edge; (b): the chain starts 1.1 Å from the zigzag edge; (c): the chain starts 2.9 Å from the armchair edge.

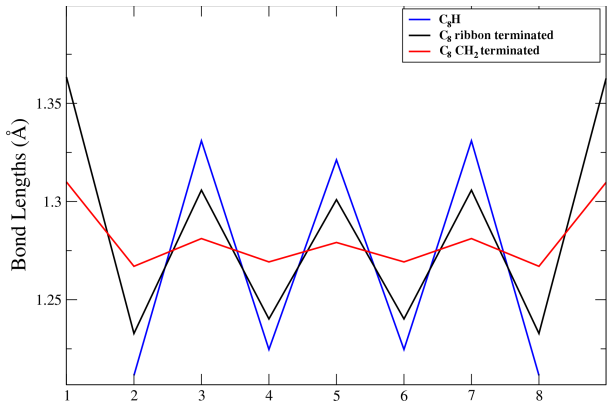

**Figure 3** Individual bond length in a $C_8$ chain with different terminations.

We made several simulations for systems of this kind, including chains from 4 to 14 atoms, with full geometrical relaxation of the structures. Figure 1 shows a set of typical results for the 8-atoms chain. Besides the binding energies, we also report the bond length alternation (BLA), expressed in picometers.

The BLA should in principle be a criterion to distinguish polyynes (with single and triple bonds) from cumulenes (with double bonds). Naively, one would expect that $sp^3$-terminated chains are polyynes, while $sp^2$-terminated ones are cumulenes. As we demonstrate below, the situation is quite more complex and the nature of the bonding along the chains is non trivial.

Recently, nice experimental observation by transmission electron microscopy [11, 12] have shown such linear carbon chains attached to graphene edges. The experimental geometries, as shown in figures 1 and 4 of Ref. [12], are very similar to those predicted to be the most stable ones by our DFT calculations, e.g. the straight chain between two zigzag edges and the "handle" structures realized by bent chains. However, present TEM images cannot discriminate between double bonds and single/triple ones, hence theoretical results are necessary to understand if the chains have a polyynic or cumulenic character.

Considering again our prototypical 8-atoms chain attached to the zigzag edges of graphitic nanofragments, we show in Fig. 3 the computed bondlenghts along the chain (black), compared with those of a pure polyyne (blue) and a pure cumulene (red). The alternation is clearly intermediate between the two limiting cases: the chains attached to a graphene edge have hence a mixed polyyynic-cumulenic character.

Another intriguing aspect of carbynes terminated on an edge of an $sp^2$ fragment comes from the breaking of the chain axial symmetry, since $sp^2$ termination defines a specific termination plane.

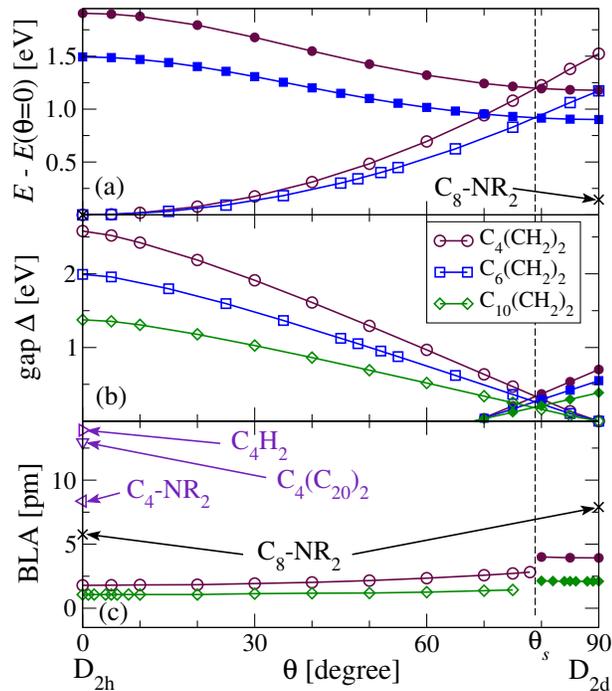

**Figure 4** Total torsional energy (a), Kohn-Sham electronic gap (b), and bond-length alternation (c) as a function of the twist angle $\theta$ for representative even-numbered sp-carbon chains with different terminations. Open/filled symbols refer to the low/high-spin electronic configurations.

This point can be clearly illustrated in the case of a purely cumulenic chain. It is well known that the equilibrium geometry of hydrogen-terminated cumulenes ($C_nH_4$) is either planar or staggered, according to the number of carbons being even or odd. This is simply due to the alternating orientation of $\pi$-bonds along the chain. In fact, long cumulenes turn out to be easy to bend, but they are rotationally stiffer.

We can hence expect that $sp^3$-terminated chains are free to rotate around their axis, and can be classified as polyynic, while chains which are $sp^2$-terminated on a graphitic edge, instead, are affected by the relative angular orientation of their terminal groups.

In a disordered material a large fraction of the edge-terminated chains will probably be torsionally strained.

**3.2 Effects of torsional strain** Figure 4 shows the result af explicit calculations of the effects of torsional strain, performed on a representative set of our optimized structures. The latter include pure cumulenes, chains terminated on a double phenilic unit, and chains connecting nanoribbon edges and fullerenic cages. We report total energies, HOMO-LUMO gaps, and bond-length alternation

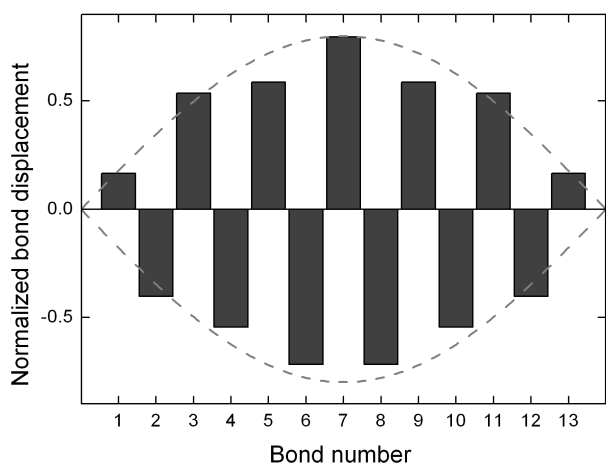

**Figure 5** The pattern of longitudinal atomic displacements of the Raman active $\alpha$ mode of an untwisted $C_{14}$ chain.

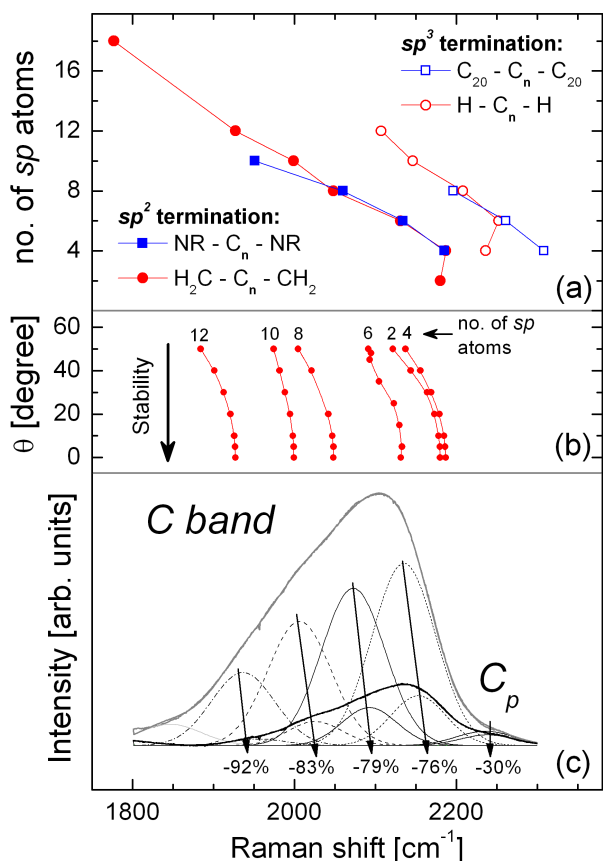

**Figure 6** (a) The computed frequency of the Raman $\alpha$-mode (horizontal scale for carbon chains of different length $n$ (vertical scale) and with different terminations. (b) The softening of this mode for $CH_2$-terminated chains as a function of twist angle $\theta$ (vertical scale). (c) Experimental Raman spectra for pristine cluster-assembled sp-sp$^2$ film (grey line) and for the same material after 2 days exposure to He, 100 Torr (black line). The underlying Gaussians report the empirical analysis of both spectra, resulting in 5 components at frequencies separated by approximately 80 cm$^{-1}$. Individual components display different decay rates, beside becoming narrower and undergoing a $\sim$ 10 cm$^{-1}$ blueshift.

as a function of the twist angle for a set of even-numbered chains, having then their ground-state geometry in the planar configuration. Full symbols refer to spin polarized configurations, while empty symbols are used for non spin-polarized ones. By forcibly twisting the chain termination, and letting all other degrees of freedom to relax, the total energy undergoes a substantial increase, which is larger for shorter chains and for chains having a more pronounced cumulenic character. The total energy increase is accompanied by a small increase of the BLA and by a substantial reduction of the HOMO-LUMO gap, which would become exactly zero in the $D_{2d}$ configuration due to the degeneracy of the $\pi$-bonds oriented in the two orthogonal planes containing the molecular axis. Moreover, beyond a certain angle the system undergoes a magnetic instability, switching to the spin-polarized configuration (which for even numbered chains is favored in the $D_{2d}$ geometry).

**3.3 Raman spectra** In the light of these results, we expect chain termination details and geometry to influence the vibrational properties of the chains. We focus, in particular, on their Raman signature, associated with the high-frequency $sp$ C-C stretching modes, and perform ab-initio phonon calculations for a set of our optimized structures. We concentrate on the so called "alpha mode", which is known to possess the strongest Raman activity and has a typical displacement pattern as shown in Fig. 5. Since the eigenvector of this mode goes to zero on the chain ends, its frequency does not depend on the mass of the chain termination, and in fact we verified that no changes appear by imposing fixed or free boundary conditions. In fact, the frequency of the alpha mode clearly depends on three factors: the chain lenght, the type of termination ($sp^2$ or $sp^3$), and, for $sp^2$ termination, the relative orientation of the terminating planes.

Figure 6a shows the frequency of the alpha mode as a function of the chain lenght. A first remark is that at

*fixed lenght* the chains with an $sp^2$ termination are systematically redshifted by about 50 cm$^{-1}$, with respect to $sp^3$-termination. From Fig. 6a one can also conclude that any spectral feature above 2200 cm$^{-1}$ must come from $sp^3$-terminated chains, since the highest mode of $sp^2$-terminated chains, that is found in the 4-atoms chain, remains below this value. Figure 6b focuses on the effects of the torsional strain: the frequency of the alpha-mode of $sp^2$-terminated chains is shown as a function of the twist angle. The Raman mode shows a clear *softening* in the torsionally strained configurations. One can hence predict that a system containing a mixture of strained and nonstrained chains will display broadened peaks, each associated with one family of $sp^2$-terminated chains.

The picture coming out from the calculations turns out to be consistent with the experimental findings of in-situ Raman characterization of cluster-assembled $sp - sp^2$ carbon films [4,5]. Figure 6c displays the spectra acquired with the 514.5 nm wavelength of a Ar laser[2] on the pristine samples and after 2 days' exposure to He (100 Torr). Several components can be identified in the spectra, behaving consistently with the attribution to $sp$ carbon chains of different lengths. Indeed individual components are characterized, on one side, by faster decays of peaks at lower Raman shift (i.e. longer chains are more reactive), on the other side the redshift and the narrowing of each individual component is consistent with the expected faster decay of the more strained chains (which are also more reactive, as illustrated in Fig. 4a). Remarkably, the absence of shift observed for the highest frequency peak is consistent with the fact that no torsional strain effects applies to the $sp^3$-terminated chains, which can rotate freely around their axis.

Moreover, the presence of strongly strained, and hence small-gapped chains (here an odd-numbered chain forced to stay in the planar geometry) can explain the higher electrical conductance of the samples immediately after deposition and is consistent with the observed decay. The conductance properties of carbyne-bridged $sp^2$ structures are being studied actively[13–15].

**4 Conclusions** In conclusion, we have seen that carbon linear chains can be substantially stabilized by $sp^2$ termination on the edge of graphitic fragments: this finding is in nice agreement with the structures recently detected by transmission electron microscopy. The nature of the $sp$ bonding along the chain is nontrivial, since in most cases C-C bonds cannot be classified as either purely double or a pure sequence of single-triple bonds. Moreover, when the chains are $sp^2$-terminated, a memory of the orientation of the termination plane propagates along the chain, through the sequence of $\pi$-bonds, making the system sensitive to torsional strain. The theoretical picture coming out from the calculations is consistent with the experimental Raman spectra measured in cluster-deposited carbon films, and with the observed time decay of the $sp$-related Raman peaks.

Our calculations also suggest that several other properties of $sp^2$-terminated carbon chains are likely to be influenced by torsional strain, e.g. electrical conductivity, optical spectra, and magnetic properties.

**Acknowledgements** The research leading to these results has received funding from the European Community's Seventh Framework Programme (FP7/2007-2013) under grant agreement No. 211956 (ETSF-i3). We acknowledge generous supercomputing support from CILEA.

---

[2] Similar spectra were obtained with the 488 nm excitation Ar laser, and reported in previously published work [6].